\documentclass{article}

\usepackage[preprint]{neurips_2023}

\usepackage[utf8]{inputenc} 
\usepackage[T1]{fontenc}    
\usepackage{hyperref}       
\usepackage{url}            
\usepackage{booktabs}       
\usepackage{amsfonts}       
\usepackage{nicefrac}       
\usepackage{microtype}      
\usepackage{xcolor}         
\usepackage{amsmath}
\usepackage{graphicx} 

\title{Generative Modeling of Bach-Style Symbolic Music: A Comparative Study of Autoregressive, Latent-Variable, and Adversarial Approaches}

\author{%
  Dezhi Yu\textsuperscript{*} \\
  \texttt{dezhiy@stanford.edu} \\
  \And
  Kyuil Lee\textsuperscript{*} \\
  \texttt{kyuil@stanford.edu} \\
  \And
  Yongkang Huang\textsuperscript{*} \\
  \texttt{yka@stanford.edu} \\
}

\begin{document}

\maketitle
\begingroup
\renewcommand{\thefootnote}{*}
\footnotetext{These authors contributed equally to this work.}
\endgroup

\begin{abstract}
We study generative modeling of Bach-style symbolic piano music using a shared MIDI corpus and three model families: autoregressive LSTMs with attention, latent-variable models including recurrent VAEs and vector-quantized VAEs, and generative adversarial networks. We compare their ability to model polyphonic note sequences, learn useful latent representations, and generate stylistically coherent compositions. Our experiments show that the autoregressive LSTM with attention produces the most musically coherent samples, while vector quantization helps mitigate posterior collapse and yields more structured outputs than conventional recurrent VAEs. The adversarial approach captures local pitch patterns but remains difficult to train and generalizes less reliably to Bach's style. These results highlight the relative strengths and failure modes of autoregressive, latent-variable, and adversarial approaches for symbolic music generation.
\end{abstract}

\section{Our code}
\url{https://github.com/cs236-bach/cs236_bach}

\section{Introduction, Motivation, Related Work}
In the classical music realm, Johann Sebastian Bach’s works stand as exemplars of structured elegance and emotional depth, showcasing complex patterns within melodious arrangements. His consistent use of Baroque techniques, particularly counterpoint, establishes a unique, enduring style that continues to enchant audiences. Enter "BachBot", a project blending the historical richness of Bach’s music with Generative AI’s futuristic potential. Our goal is to explore various generative models in their abilities to generate symbolic music in Bach's style. Specifically, we explore using LSTMs with attention, VAEs, and GANs to generate and learn hidden representations for Bach's music.

\section{Related Work}
In the domain of music composition, many efforts have been put into developing models based on Long Short-Term Memory (LSTM) due to their ability to handle time-series data effectively. One notable work by Kong et al., \cite{kong2021bach}, involved fine-tuning an LSTM model to generate music in the style of Bach. Recently, there's also been a rise in the use of deep generative models like Variational Autoencoder (VAE) for music generation. For example, the method mentioned in \cite{DBLP:journals/corr/abs-2102-05749} used segments from audio files for self-supervised learning of the VQ-VAE, with an additional style encoder to handle different music styles. Another of note is the MusicVAE \cite{roberts2019hierarchical}, which uses hierarchical decoders to generate music. \cite{copet2023simple} introduced a model that uses an autoregressive transformer-based decoder, which can be conditioned on text or melody inputs to control the music generation process.

\section{Dataset}
We scraped all of Bach’s known piano works in the MIDI format from \cite{CompleteBachMidiIndex}. It includes a wide array of Bach's compositions, such as the Well-Tempered Clavier (both parts I and II), the Art of Fugue, Two-part Inventions, Goldberg Variations, and more.

\subsection{Data Pre-Processing}
MIDI files encode musical information such as notes, durations, and velocity. We transformed the MIDI files into a sequence of multihot vectors of dimension 88 (one for each possible piano key), where a value of 1 represented that the note was being played at that specific time step. If no 1's exist in the vector, then this signifies a rest. We left out a lot of other information such as the velocity of each note (how loud each note is) to simplify the problem for our models. We used \cite{RaffelEllis2014} to interact with the MIDI files.

The time step that each multihot vector represents is determined by the shortest duration of notes that makes up more than 20 percent of a piece. For example, if sixteenth notes are the shortest notes that appear in more than 20 percent of the piece, then it becomes the default time step. Similarly, if a piece has barely any sixteenth notes, but many 8th notes, then the duration of the 8th notes becomes the default time step of the piece. We do this because this allows the models to not have to constantly deal with held notes and focus on the variational patterns in the music. We also explicitly filter out long pauses from our dataset as well.

\section{Technical Approach}
\subsection{Baseline: Autoregressive Long Short-Term Memory}
We first do a baseline experiment using LSTMs. LSTMs are well-suited for processing longer sequences due to their ability to maintain long-term dependencies, making them useful for music generation.

\begin{figure}[h]
    \centering
    \includegraphics[width=0.5\textwidth]{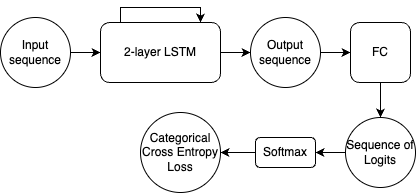}
    \caption{LSTM architecture}
    \label{fig:vae}
\end{figure}

We have a 2-layer LSTM with a hidden state size of 512 and a dropout rate of 0.5. The LSTM generates a sequence of length 32, which is then passed through a single FC layer to generate the sequence of logits. The logits are passed through a softmax function to calculate the probability distribution of each note in the sequence. Using the sequence of categorical distributions, we calculate the negative log-likelihood of the input sequence by calculating the softmax loss (categorical cross entropy) for each element in the sequence and then adding up the results.

\begin{equation*}
log p_\theta (x) = -\sum_{s = 1}^{32}\sum_{i=1}^{88} x_{s, i} \log(\hat{x}_{s, i})
\end{equation*}

$x_{s, i}$ is the $i^{th}$ element of the $s^{th}$ note in the sequence, and $\hat{x}_{s, i}$ is the categorical distribution the model predicts for this note. Since $x$ is one-hot encoded, only one out of 88 values of $x_{s, i}$ will be set to 1 and the rest to 0.

We use the ADAM optimizer with a learning rate of 0.001.
\subsection{Using Recurrent VAEs}

We also try to learn a latent structure for Bach’s music by training with a Variational Autoencoder. Here, we modify a classical VAE by making it handle recurrent data, using LSTMs.

\begin{figure}[h]
    \centering
    \includegraphics[width=0.5\textwidth]{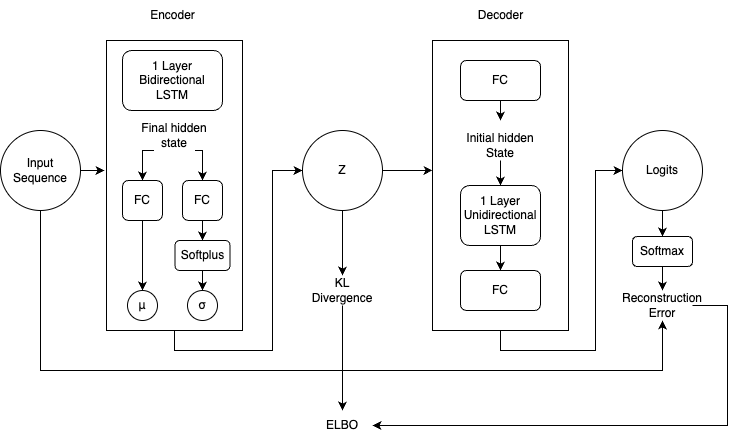}
    \caption{Baseline VAE Architecture}
    \label{fig:vae}
\end{figure}

The encoder is made recurrent with a 1-layer bidirectional LSTM with a hidden state size of 512. Making the LSTM bidirectional allows it to learn patterns in both directions of the sequence, leading to greater flexibility. The final hidden state of the LSTM is sent through two separate neural networks, which calculate $\mu$ and $\sigma$. 

$\mu$ and $\sigma$ are then used to sample the latent variable $z$, using the standard reparametrization trick. 

The latent variable z is sent to the decoder, where it goes through an FC layer to generate the first hidden state for the bottom-most layer of a 1-layer (unidirectional) LSTM. The LSTM generates a sequence of 32 values, which go through a shared FC layer to generate the sequence of output logits. We use softmax on the logits to calculate the categorical distribution for each note in the sequence and calculate the log-likelihood similar to our baseline example.

\begin{equation*}
log p_\theta (x|z) = -\sum_{s = 1}^{32}\sum_{i=1}^{88} x_{s,i} \log(\hat{x}_{s, i})
\end{equation*}

We approximate the NELBO term 
\begin{equation*}
    \mathrm{NELBO}= -\mathbb{E}_{q_\phi(z|x)}\left[\log p_\theta (x | z) \right] + \mathrm{D}_{KL}(q_\phi (z | x) || p(z)) 
\end{equation*}
with Monte Carlo, using the sampling process described above, and we try to minimize the NELBO. We also use the ADAM optimizer with a learning rate of 0.001.

\subsection{Hierarchical Recurrent VAEs}
The above approach unfortunately results in posterior collapse, as we will talk more about in in the Results section. To address this issue, we also try an architecture suggested by \cite{roberts2019hierarchical} to make the decoder hierarchical, dividing the original decoder into two layers of LSTMs, where the top layer is called the conductor and the bottom layer is the decoder.

\begin{figure}[h]
    \centering
    \includegraphics[width=0.5\textwidth]{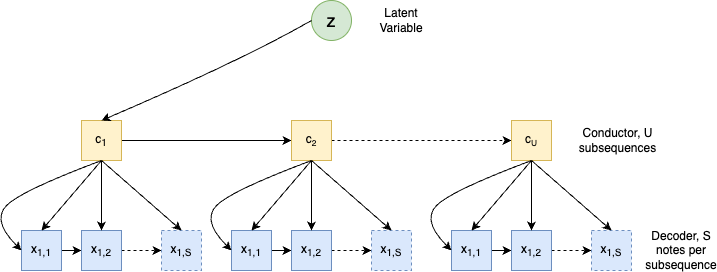}
    \caption{Hierarchical Decoder in VAE}
    \label{fig:vae}
\end{figure}

The output sequence is subdivided into $U$ subsequences. The conductor first generates embeddings for each of the subsequences, $c_1, c_2$, …, $c_U$. Then, each embedding is used by the lower-level decoder to generate $S$ notes per subsequence, where $S * U = N$, $N$ is the total number of input/output notes. The embedding is both passed in as the initial input into the lower-level decoder and also concatenated into the hidden states of the decoder by attaching them to the previous output of the decoder. 

Breaking up the generation such that the decoder only handles a smaller, fixed number of notes, while taking in the embedding vector from the conductor into the input and hidden states, forces the decoder to depend more on $z$ as the sole context in note generation (whereas in the previous example, the effect of $z$ gradually declined as more notes were produced in the output).

\subsection{Vector Quantized VAEs}
Finally, we also experimented with vector quantized VAEs (VQVAEs) as another way to avoid the posterior collapse problem \cite{vandenOord2017NeuralDiscrete}. At a high level, the VQVAE learns a set of discrete embedding vectors to represent the latent space (called the codebook). In our case, we divided the input sequence into groups of 4 notes, which the model then learned to represent as a discrete embedding vector - basically learning 4-note patterns. Once this embedding was learned, an autoregressive model (LSTM) was trained on top of the embedding space, in order to learn how to piece the embedding vectors together (basically, piecing together four-note patterns). Then, the embeddings were sent to the trained decoder from the VQVAE to convert them to actual notes. An inspiration was drawn from the Bachsformer \cite{Melucci2022Bachsformer}, but we used different encoder and decoder models as well as a different autoregressive model in our experiments.

\begin{figure}[ht]
    \centering
    \begin{minipage}{0.45\textwidth}
        \centering
        \includegraphics[width=\textwidth]{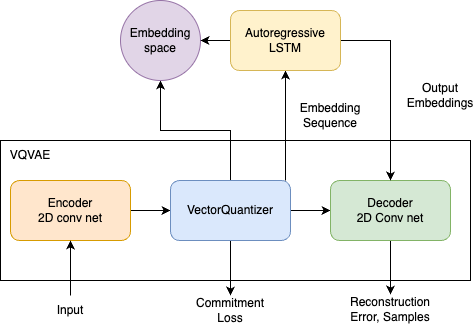} 
        \caption{VQVAE Architecture}
        \label{fig:figure1}
    \end{minipage}\hfill
    \begin{minipage}{0.45\textwidth}
        \centering
        \includegraphics[width=\textwidth]{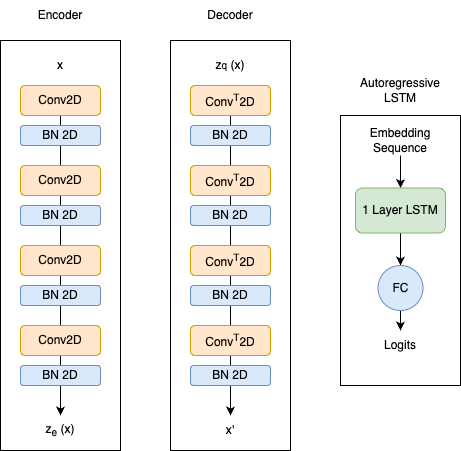} 
        \caption{VQVAE - Encoder, Decoder, Autoregressive LSTM}
        \label{fig:figure2}
    \end{minipage}
\end{figure}

For the VQVAE, the data had to be organized slightly differently. The data was initially organized as $(batch, seq_len=256, 88)$, which was converted to $(batch, 64, 4, 88)$ - a sequence of 4-note patterns. For the VQVAE to learn the codebook, this was modified to be $(batch * 64, 4, 88)$ to only focus on the four note patterns. 

Once the VQVAE was trained, we used its encoder and vector quantizer to generate a sequence dataset for our autoregressive LSTM, by feeding in our input sequences of shape $(batch, 64, 4, 88)$ to get a sequence of embedding vectors of shape $(batch, 64, embedding_dim)$. The sequence of 64 vectors was broken up into a sequence of 5 vectors through a sliding window method, where 4 vectors would be used as an input sequence and the 5th vector would be the target vector to be predicted during training. This resulted in $64 - 5 + 1 = 60$ input-output pairs for our autoregressive model.

Once the autoregressive LSTM is trained on this generated dataset, its output was then sent to VQVAE’s decoder to generate the actual notes.

\subsection{VQVAEs - more details}

The VQVAE has 4 parts - the encoder, the vector quantizer, the decoder, and an autoregressive model that learns over the quantized latent space $z$.

The encoder first maps the input variable $x$ into a continuous latent variable $z_e$. Then, the vector quantizer converts $z_e$ to a quantized embedding space $e$ by mapping to the nearest embedding vector $e_i$, denoted $z_q(x)$ (where distance is calculated by Euclidean distance). Finally, the embedding vector $z_q(x)$ is sent to the decoder, which tries to reconstruct the input $x$.

The loss is given by the following:

$$
L = \log p(x | z_q (x)) + \beta||z_e(x) - \mathrm{sg}[e]||_2^2
$$

The first term is the usual reconstruction loss, and the second term is called the “commitment” term, which forces the encoder to output something similar to an embedding vector and to prevent the encoded vector from growing too much. The $e$ is simply $z_q(x)$, but denoted as $e$ since its gradient is not propagated, denoted by “sg” for “stop gradient”.

The embedding space itself is learned by calculating the exponential moving averages of the encoded inputs. Basically, the algorithm is similar to the K-means algorithm, where the embedding vector becomes the average of the encoded input vectors that are closest to the embedding vector. However, since we’re dealing with one batch at a time, we calculate the embedding vector using exponential moving averages, as detailed in the appendix of the VQVAE paper.

\subsection{Autoregressive LSTM with Attention}
Our project adopts an autoregressive LSTM model with an attention mechanism. This approach is designed to leverage the LSTM's sequential data processing capabilities while the attention mechanism provides a dynamic focus on relevant parts of the input sequence.

\begin{figure}[h]
    \centering
    \includegraphics[width=0.7\textwidth]{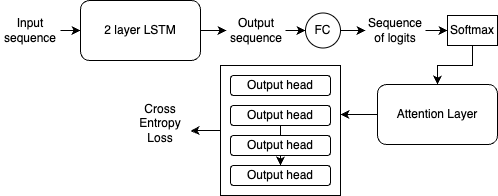}
    \caption{LSTM with Attention Architecture}
    \label{fig:vae}
\end{figure}

The model's architecture can be summarized as follows:

\begin{itemize}
    \item Input Layer: Represents multi-dimensional musical data.
    \item LSTM Layers: A sequence of 2 LSTM layers.
    \item Attention Layer: Processes LSTM outputs to create a context vector, focusing on crucial sequence parts.
    \item Multiple parallel output heads, each predicting different voices of the musical piece. Each head comprises linear layers, ReLU activation, dropout for regularization, and a final linear layer mapping to output sizes.
    \item Outputs: The notes generated by each head are combined into a sequence of multihot output vectors, representing the actual music.
\end{itemize}

Mathematically, the LSTM model with attention can be expressed as:

$$\operatorname{Attention}(Q, K, V)=\operatorname{softmax}\left(\frac{Q K^T}{\sqrt{d_k}}\right) V$$

where $Q, K$, and $V$ represent queries, keys, and values, respectively, and $d_k$ is the scaling factor derived from the dimensionality of the keys.

We use cross-entropy loss used in sequence generation tasks:

$$
L(\theta)=-\sum_{t=1}^T \sum_{i = 1}^{88} y_{t,i} \log \left(\hat{y}_{t,i}\right) 
$$

where $y_{t,i}$ denotes whether the $i^{th}$ pitch at time $t$ is played and $\hat{y}_{t,i}$ is the predicted probability of the note being played by the model. 

In selecting this algorithm, we aimed to address the limitations of traditional LSTMs in handling long sequences where certain notes may be more influential than others. The attention mechanism's introduction is a strategic decision to amplify the model's focus on such influential notes, thereby generating a composition that not only flows logically but also resonates with the stylistic nuances of the genre.

\subsection{Generative Adversarial Networks}
For audio synthesis, we adopt the following architecture inspired by \cite{Yu2020ConditionalLF}, after trying various combinations of discriminator and generator.

\begin{figure}[h]
    \centering
    \includegraphics[width=0.5\textwidth]{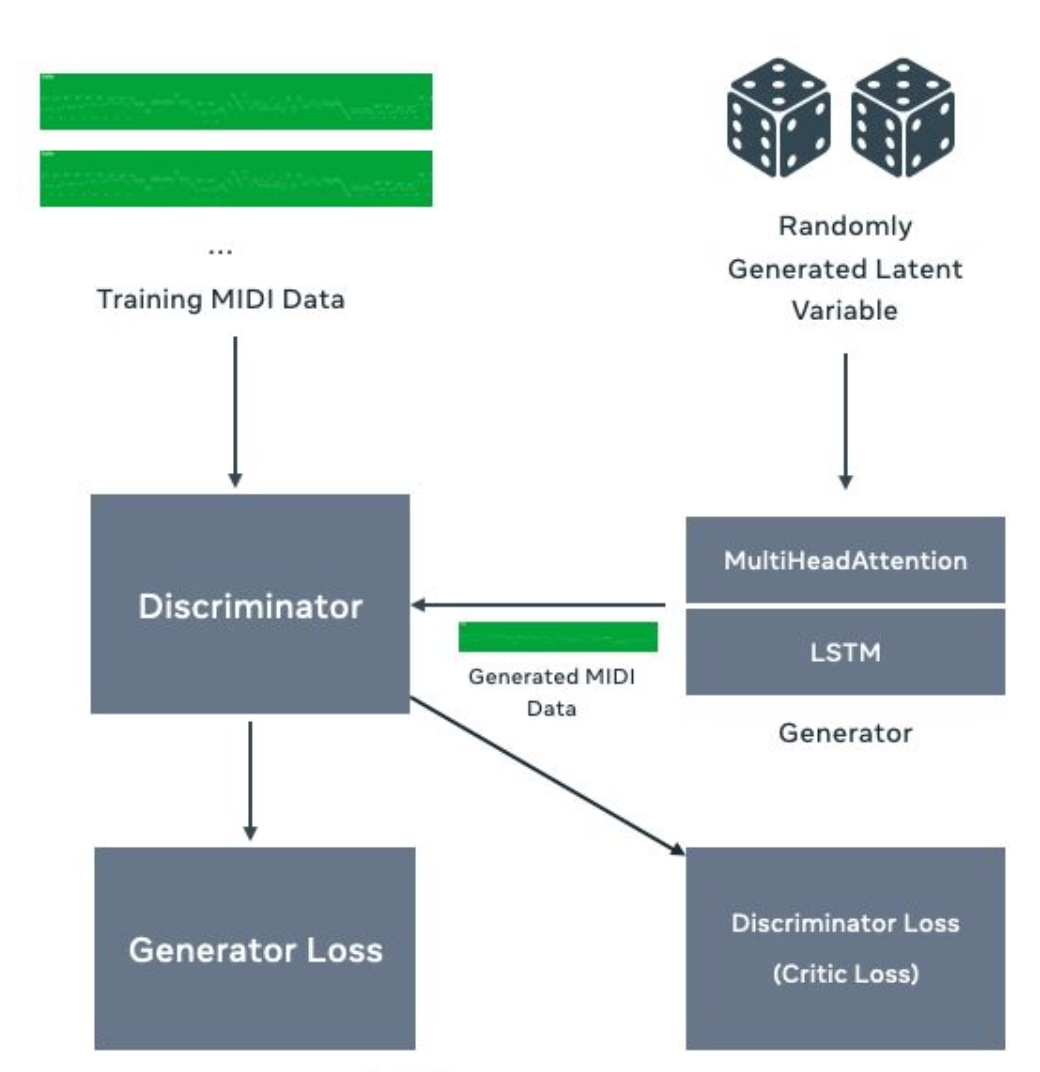}
    \caption{LSTM with Attention Architecture}
    \label{fig:vae}
\end{figure}

The input is the vectorized MIDI file that has the shape of ${[batch, sequence\_length, 88]}$, similar to before. We treat the batch size and sequence length as hyperparameters, as both can significantly impact GAN training. A larger batch size is preferable to reduce noise during training updates, enhancing the effectiveness of the GAN training process.

Our generator is based on a combination of LSTM and MultiHeadAttention layers. The input for the generative model is a latent variable vector sampled from $N(0; I)$. The latent variable dimension is also deemed as a hyperparameter. 

During the training process, generally, the discriminator is relatively easier to train compared to the generator, such that even a linear neural network setup is capable of matching the learning efficacy of the generator. In the final set-up, we adopt a similar set-up for the generator as a discriminator to use the LSTM and attention model combination in order to keep the consistency of model capability and simplify hyperparameter tuning.

We tried using the default GAN minimax loss, which had a tendency to train the discriminator faster than the generator as well as vanishing gradients, so we instead ended up using the Wasserstein GAN \cite{arjovsky2017wasserstein} loss function during our training.

$$
L_{\text{C}} = \mathbb{E}_{\mathbf{x} \sim \mathbb{P}_{data}} [f(\mathbf{x})] - \mathbb{E}_{\tilde{\mathbf{x}} \sim \mathbb{P}_\theta} [f(\tilde{\mathbf{x}})]
$$

$$
L_{\text{G}} = -\mathbb{E}_{\tilde{\mathbf{x}} \sim \mathbb{P}_{\theta}} [f(\tilde{\mathbf{x}})]
$$

However, while WGANs generally reduce the risk of gradient vanishing, they can still encounter issues with gradient exploding, In the original formulation of WGANs, weight clipping is used to enforce the Lipschitz constraint on the critic (or discriminator). There are also other variations of WGAN loss to introduce Lipschitz constraint like WGAN-GP \cite{gulrajani2017improved}

\section{Results \& Analysis}
Since we experimented with a number of different models (autoregressive models, latent variable models (VAEs), and GANs), each had its own experimental procedures as well as metrics. We will go into our results and analysis for each section separately.

\subsection{LSTM Baseline}
We noticed that the probability distribution for the notes often collapses to repeat the last note in the sequence, resulting in one prolonged note for the samples. The second most probable state was the rest. This phenomenon started happening as early as 300 steps of training, which was before the model had learned any meaningful structure behind Bach’s music.

So instead, we modified our approach by adding a dropout layer, as well as sampling by ignoring the top two most probable notes (repeated and rest). After 10,000 steps of training, we started seeing a back-and-forth between two notes (C, D, C, D, …), but also sometimes seemed to a semblance of reasonable music, with partial scales and arpeggios. 

Overall, we saw that the baseline model focused more on the micro-patterns that exist in the music, such as tremolos or short scales, but failed to capture Bach's overall style.

\subsection{VAE}
For the initial two VAE models (single layer decoder, hierarchical decoders), the data was organized into shapes [batch, 32, 88], representing sequences of 32 piano notes. For the VAE with the hierarchical decoder, we used subsequence lengths of 8, where there were 4 notes per subsequence. We experimented with different latent space dimensions (32, 64, 128, 256, 512) for each model. We measured the reconstruction error and the KL divergence for these models as our metric.

\begin{figure}[ht]
    \centering
    \begin{minipage}{0.5\textwidth}
        \centering
        \includegraphics[width=\textwidth]{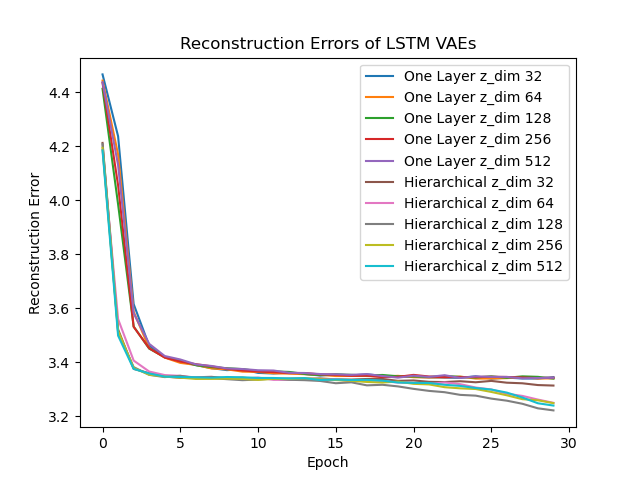} 
        \caption{Reconstruction Errors}
        \label{fig:figure1}
    \end{minipage}\hfill
    \begin{minipage}{0.5\textwidth}
        \centering
        \includegraphics[width=\textwidth]{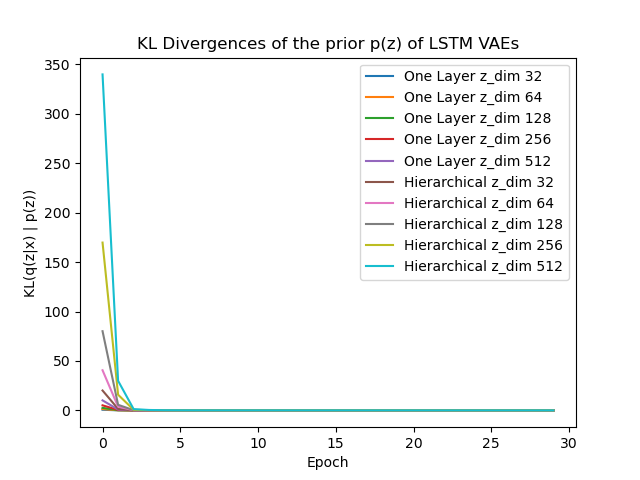} 
        \caption{KL Divergences}
        \label{fig:figure2}
    \end{minipage}
\end{figure}

As you can see, both approaches result in quick posterior collapses, but the hierarchical decoder models fall into collapse slightly later than the single-layer models, resulting in lower reconstruction error.

The different latent space dimensions turned out to not affect the training much. We tried several methods, such as teacher forcing, annealing the KL divergence term, scheduling the learning rate, and more, but they did not change the result much. It seems that a fundamental shift to the design of the model is needed. Since the reconstruction error was relatively high, the resulting music was not able to capture Bach’s style. The output was not completely random, in that it sounded like music, with reasonable note intervals, and some scale / arpeggio-like sequences, but it sounded more like jazz than Bach.

\begin{figure}[h]
    \centering
    \includegraphics[width=0.5\textwidth]{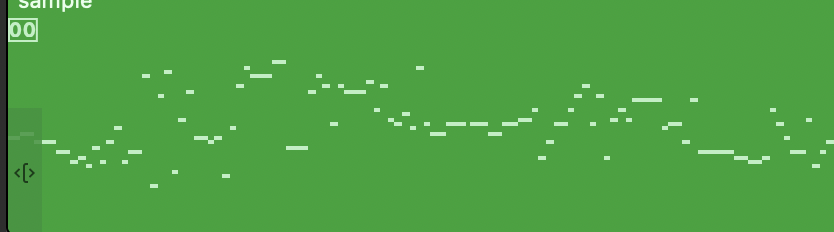}
    \caption{MIDI generated by hierarchical VAE}
    \label{fig:vae}
\end{figure}

For the VQVAE, as explained in an earlier section, the input was organized into the shape [batch, 64, 4, 88]. We experimented by varying the size of the embedding space (32, 64, 128), and measured the reconstruction error.

\begin{figure}[ht]
    \centering
    \begin{minipage}{0.5\textwidth}
        \centering
        \includegraphics[width=\textwidth]{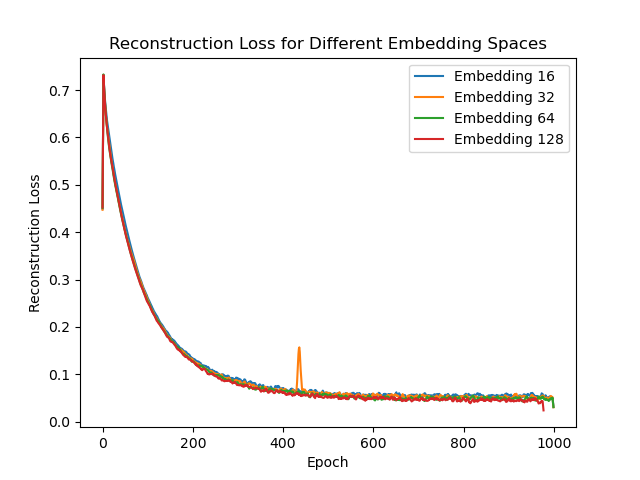} 
        \caption{Reconstruction errors for VQVAE}
        \label{fig:figure1}
    \end{minipage}\hfill
    \begin{minipage}{0.5\textwidth}
        \centering
        \includegraphics[width=\textwidth]{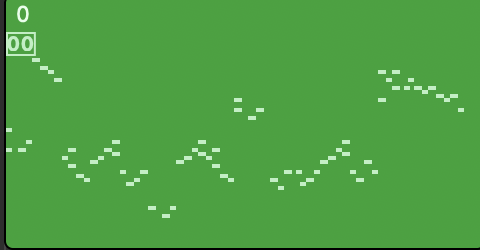} 
        \caption{MIDI generated by VQVAE}
        \label{fig:figure2}
    \end{minipage}
\end{figure}

As you can see (if you look closely), the reconstruction loss drops further when you increase the size of the embedding space, which makes sense since this allows the model to learn more diverse 4-note patterns, allowing it to output a more flexible note sequence.

The note patterns exhibit clearer patterns, such as going up and down, and having scale-like sequences. The output audio was also able to capture a semblance to the Baroque style, showing that using VQVAEs is a viable strategy.

\subsection{LSTM with Attention}
We split the data into training and testing data in a 90/10 ratio for validation after training. 

The window size for the input data is set as a hyperparameter, training the model with context windows for each prediction. The skip steps parameter is fixed at 16. The dropout rate of 0.2 in the LSTM layers, coupled with an additional 0.5 dropout in the task heads, effectively counters overfitting, making the model more generalizable.

\begin{figure}[h]
    \centering
    \includegraphics[width=0.95\textwidth]{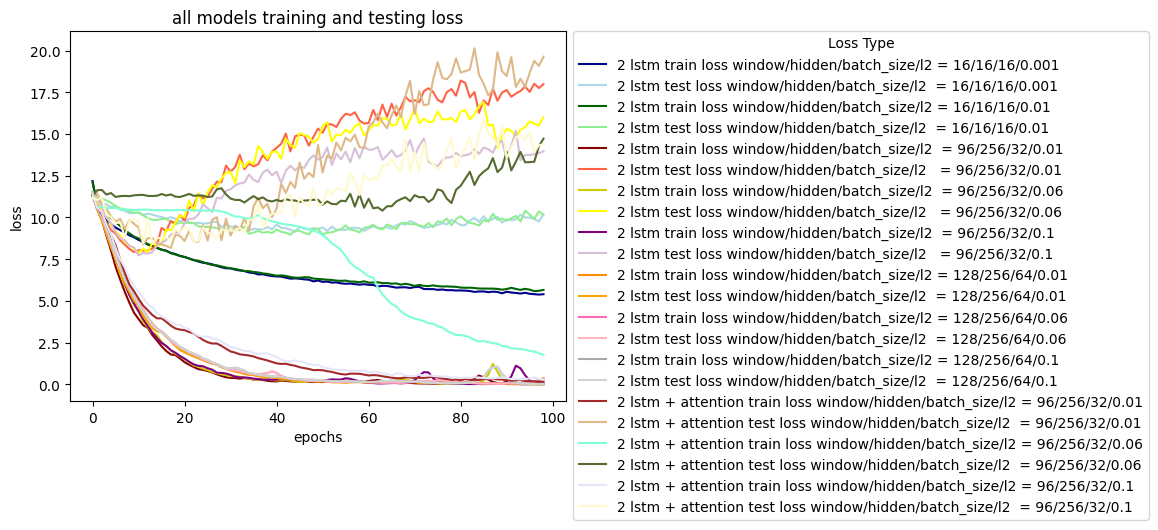}
    \caption{LSTM with Attention, Results}
    \label{fig:vae}
\end{figure}

The chart displays training and testing loss across epochs for various LSTM models. The loss is a measurement of the model's prediction error, with lower values signifying more accurate predictions.

The baseline model, utilizing two LSTM layers with minimal hyperparameters (window size, hidden size, batch size, L2 = 16, 16, 16, 0.001), starts with a training loss of approximately 12.4. Over 100 epochs, it converges to a training loss of around 6.3, indicating a decent learning curve but suggesting there is room for improvement.

Drawing on insights from the article \cite{longsongIndex}, it's understood that a two-layer LSTM constitutes an effective model structure. These articles \cite{lstmmusicIndex} \cite{bahdanau2016neural} reveal that incorporating an attention layer can enhance the quality of the generated music. The proposed model with an enhanced configuration (two LSTM layers plus attention, window size, hidden size, batch size, L2 = 96, 256, 32, 0.06) shows superior performance with its training loss nearing zero after 100 epochs. This suggests a highly effective learning process and a model that fits the training data extremely well but also means that the model can easily overfit.

\begin{figure}[h]
    \centering
    \includegraphics[width=0.95\textwidth]{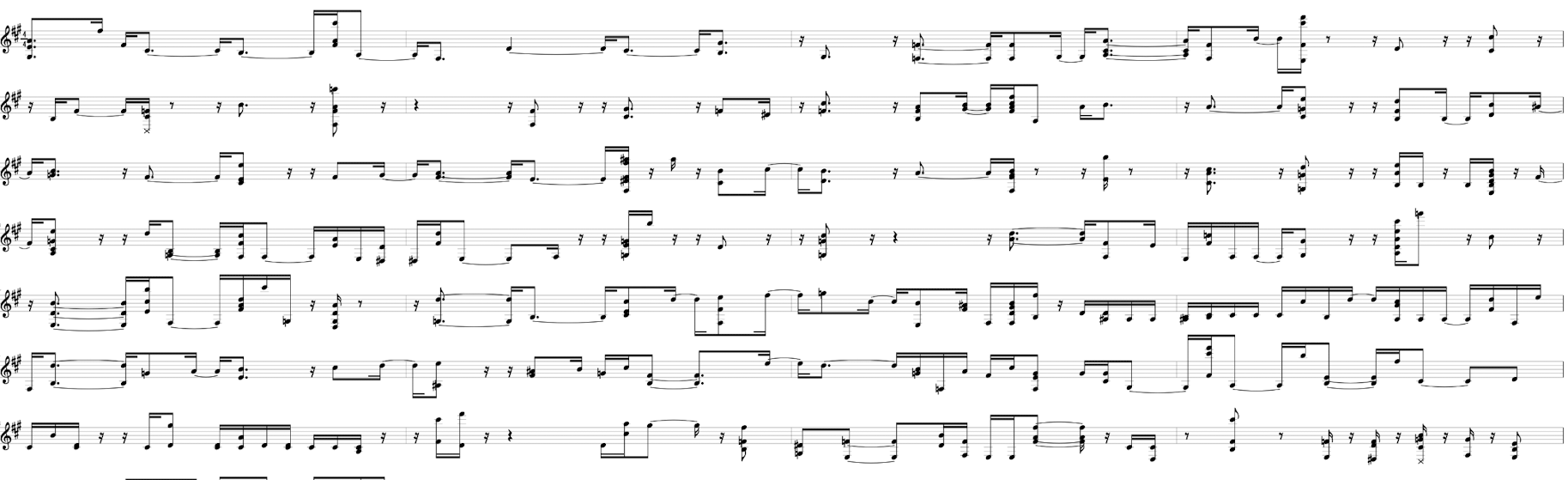}
    \caption{LSTM with Attention, Results}
    \label{fig:vae}
\end{figure}

The above image represents a score of the generated Bach-styled music. As you can see, the chords and melodic progressions are quite sensical and reminiscent of a Baroque style. From listening to the music, we found that the generated music has a strong semblance to Bach’s music, and this matched our intuition that solving the problem autoregressively was a much easier problem than learning latent representations or training adversarially, resulting in the best output.

\subsection{Generative Adversarial Networks}

For GAN training we use the 64 latent variables for the generator’s input. The batch size is tuned to 1024 corresponding to the input and output sequence, limited by the training hardware limitation, and majorly bottlenecked by GPU RAM.

The generator is composed of 4 hidden layers in LSTM while 8 parallel attention heads are in the discriminator. The discriminator is a simplified version of the generator with only 2 hidden layers and 4 parallel attention heads. Both the generator and discriminator have the hidden dimension set to 1024.

During the training, we adopt WGAN loss with parameter clipping to ensure the training stability, the clip threshold is set to 0.01 and -0.01. The learning rate is tuned with the generator using the learning rate of 0.0001 and the discriminator using the learning rate of 0.0002, slightly higher compared to the discriminator

With the hyperparameters mentioned above, GAN trained by 20 Epoch with both discriminator and generator loss sufficiently converged and the speed of convergence matches each other.

\begin{figure}[h]
    \centering
    \includegraphics[width=0.4\textwidth]{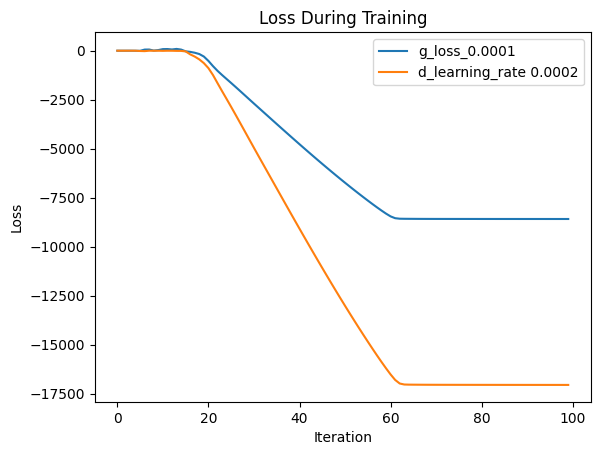}
    \caption{GAN loss during training}
    \label{fig:vae}
\end{figure}

We can observe with the intervention of parameter clipping, the loss of both discriminator and generator dropping nearly linearly.

The GAN captures 2 soundtracks analogous to the left-hand side of the piano (bass) and the right-hand side (treble). However, the music style is more like modern jazz piano with more complex harmonies and idiocracies of improvisation.

\begin{figure}[h]
    \centering
    \includegraphics[width=0.95\textwidth]{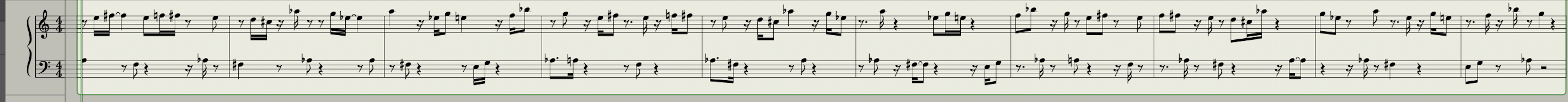}
    \caption{Music generated by GAN}
    \label{fig:vae}
\end{figure}

The outcomes of the Generative Adversarial Network (GAN) analysis indicate that the model effectively captures the pitch sequence information. However, there is significant potential for enhancement in the area of style generalization. This improvement could be achieved by incorporating additional information into the model. During training, it has been observed that the GAN does not reconstruct the training data as efficiently as Variational Autoencoders (VAE) and autoregressive-based models. In this context, introducing quantized style generalization \cite{DBLP:journals/corr/abs-1812-04948} in the loss function may prove beneficial.

Furthermore, it is recognized that the loss function used in GANs is predominantly engineering-focused. This approach to the loss function can stabilize training but may also limit the model's capabilities. Consequently, future research should also consider exploring alternative heuristics in the loss function, e.g \cite{karras2019stylebased}. Such exploration might not only enhance the stability of training but could also expand the model's potential, leading to more robust and generalizable results.

\section{Conclusion}
Overall, all of our models were able to generate music, with differing levels of similarity to Bach's style. We found that solving the problem autoregressively (LSTM with attention) gave the best results, whereas the family of VAE models and GAN models experienced their fair share of difficulties during training (posterior collapse, stability in GANs). This aligned with our belief about the difficulty of our various formulations of the problem - the autoregressive formulation being the easiest, while the adversarial method is the most difficult. Techniques such as using vector-quantized VAEs were examples of getting around these issues, and the general conclusion seems to be that we need significant architectural changes to address the challenges in these latent variable and adversarial models.

The LSTM-attention model and the VQVAE models produced compositions with a high degree of stylistic fidelity. The attention model has notably enhanced the LSTM's ability to prioritize salient features within sequences, leading to an output that is not only coherent but also creatively compelling.

Looking ahead, there are several avenues for future work to explore. One promising direction is the incorporation of a broader set of musical features such as dynamics, articulation, and expression markings. Also, it would be great to have our latent variable models learn some human-friendly structures such as scale, the emotion of the music, and more.

\newpage
\bibliographystyle{plainnat}
\bibliography{reference}
\end{document}